# *MatBase* algorithm for translating (E)MDM schemes into E-R data models

Christian Mancas[1]* and Diana Christina Mancas[2]

[1] Mathematics and Computer Science Department, Ovidius University at Constanta, Romania, christian.mancas@gmail.com

[2] Mathematics and Computer Science Department, Ovidius University at Constanta, Romania, diana.christina.mancas@gmail.com.

**Abstract:** This paper presents a pseudocode algorithm for translating (Elementary) Mathematical Data Model ((E)MDM) schemes into Entity-Relationship data models. We prove that this algorithm is linear, sound, complete, and semi-optimal. As an example, we apply this algorithm to an (E)MDM scheme for a genealogical tree sub-universe. We also provide the main additional features added to the implementation of this data science reverse engineering algorithm in *MatBase*, our intelligent knowledge and database management system prototype based on both the Entity-Relationship, (E)MDM, and Relational Data Models.

## 1. Introduction

All subuniverses of discourse are governed by business rules (e.g., ”nobody may die before birth”, ”nobody may get married before birth”, ”people’s height is between 30 and 225 cm”, etc.). A database (db) instance is useful only if it is plausible, i.e., only if its data does not violate any business rule governing the corresponding subuniverse of discourse. In dbs and software engineering business rules are formalized as constraints, i.e., closed formulas of the first-order predicate calculus with equality.

Our (Elementary) Mathematical Data Model ((E)MDM) [1] was introduced as an intermediate conceptual design level between the Entity-Relational Data Model (E-RDM) [2, 3, 4] and the Relational Data Model (RDM) [4, 5, 6]. The E-RDM includes the very powerful E-R Diagrams (E-RDs) that may be easily understood even by customers without computer science background but, despite its many extensions, it has a very limited set of constraints. The RDM has only six constraint types (domain/range, not null, default values, unique keys, foreign keys, and tuple/check). The (E)MDM has 76 constraint types, thus allowing for far more accurate conceptual data modeling and db design; they include, explicitly or implicitly, all 6 relational type ones; the non-relational type ones should be enforced by the software applications (sa) that manage the corresponding dbs.

In [4], we defined E-R data models as triples <*ERDS*, *ARS*, *ICSD*>, where *ERDS* is a set of E-RDs, *ARS* is an Associated Restriction Set, and *ICSD* is an Informal Corresponding Sub-universe Description. The associated restrictions, which correspond to the business rules governing the modeled sub-universes, are of the following five types: inclusions between object sets (e.g., *EMPLOYEES* ⊆ *PERSONS*), ranges of the attributes (e.g., *Month* between 1 and 12), compulsory (not null) attribute values (e.g., *BirthDate* compulsory), minimal uniqueness of attributes (e.g., *SSN* unique) and attribute concatenations (e.g., *Room* • *Weekday* • *StartHour* unique), and other restriction types (e.g., “no teacher may be simultaneously present in two classrooms”).

An E-RD may be single, i.e., consisting only of an entity or relationship type set and its attributes, or structural, i.e., containing any number of sets, the structural functions (i.e., binary functional relationships) relating them, and no attributes. The single ones for relationship-type sets also contain all the sets over which the relationship is defined.

*MatBase* [7, 8] is our intelligent data and knowledge base management system prototype, based on both (E)MDM, E-RDM, and RDM. It includes 3 graphic user interfaces (GUI), one for each data model, and automatically translates between them, generates corresponding sa GUIs and code for enforcing a vast majority of the (E)MDM constraint types.

In our previously published paper [9], we presented and discussed the forward *MatBase* algorithm for translating E-R data models into (E)MDM schemes. The algorithm presented in this paper is its dual, a reverse engineering type one. Its necessity arises from the following couple of reasons:

- ✓ Reflecting updates of a(n) (E)MDM schema after obtaining it from an E-R data model.
- ✓ Extracting only an E-R data model subset.

The algorithm has two optional input parameters: an object set name of the current db, and a natural radius. If the first one is not null and the radius is either null or 0, then the algorithm outputs the E-R data model of the corresponding object set (whose E-RD is a single one); for any other natural value *n* of the radius, it outputs the E-R data sub-model centered in the desired object set and having radius at most *n* (i.e., whose E-RD is a structural one having length at most *n* for at least one path starting from the desired object set); if the first parameter is null, then it outputs the whole db E-R data model (and the radius is ignored if it is not null).

After the literature review, the third Section introduces and characterizes the pseudocode algorithm used by *MatBase* to translate (E)MDM schemes into E-R data models, then presents and discusses the results of applying this algorithm to an (E)MDM scheme from [10, 11]. The paper ends with conclusions, recommendations, and a reference list.

## 2. Related work

Only *MatBase* manages (E)MDM schemes.

Quest has over 30 years of its *erwin Data Modeler* [12] success. For example, it may reverse engineer both relational (e.g., MS SQL Server, Oracle Database, IBM Db2, Sybase SQL Anywhere, SQLite) and noSQL (e.g., mongoDB, cassandra, MariaDB, neo4j) schemas into corresponding E-RDs.

Related algorithms are, however, used by all major commercial RDBMSes for translating relational schemas into E-R data models-like ones.

For example, the MS SQL Server Management Studio [13] provides a *Database Diagram* menu, part of the *Visual Database Tools*. You can create, update, and delete relational diagrams corresponding to the tables of a db. These diagrams are not "standard" E-RDs but are equivalent to them: nodes are rectangles corresponding to underlying db tables that can be thought of as single E-RDs, as their attributes are listed inside them; edges are the relations between tables, as established by the foreign keys, so that they are equivalent to structural functions and E-RDs. As the RDM is a syntactical one, no relationship-type nodes are present, all of them being of entity-type. This is consistent with our Theorem stating that the only fundamental mathematical relations for conceptual data modeling are the functions [14].

Similarly, the Oracle SQL Developer provides a *Data Modeler* tool [15].

The IBM Db2 Data Management Console [16], which is the equivalent of both MS SQL Server Management Studio and Oracle SQL Developer does not provide graphical tools. They are available from third parties, such as Dataedo [17].

Other related approaches to *MatBase* are based on business rules management (BRM) [18, 19] and their corresponding implemented systems (BRMS) and decision managers (e.g., [20 – 22]). From this perspective, (E)MDM is also a formal BRM, and *MatBase* is an automatically code generating BRMS.

3. Research Methodology

Any (*E*)*MDM schema* [1] is a quadruple $DKS = \langle S, \mathcal{M}, C; \mathcal{P} \rangle$, where $S$ is a finite non-empty *poset of sets* ordered by inclusion, $\mathcal{M}$ is a finite non-empty *set of mappings* defined on and taking values from the sets of $S$, $C$ is a finite non-empty *set of constraints* (i.e., closed Horn clauses of the first-order predicate logic with equality) over sets in $S$ and/or mappings in $\mathcal{M}$, and $\mathcal{P}$ is a finite set of *Datalog*$\neg$ *programs*, also over sets in $S$ and mappings in $\mathcal{M}$. Whenever $\mathcal{P}$ is empty, *DKS* is a *db scheme*, otherwise it is a *knowledge base one*. In the context of this paper, we are only interested in db schemes.

$(S, \subseteq)$ is a *poset of sets*, with $S = \Omega \oplus \mathcal{V} \oplus *S \oplus SysS$, where $\Omega = \mathcal{E} \oplus \mathcal{R}$ (the non-empty collection of *object sets*), where $\mathcal{E}$ is a non-empty collection of atomic *entity-type sets* (e.g., *PEOPLE*, *COUNTRIES*, *PRODUCTS*), $\mathcal{R}$ is a collection of *relationship-type sets* (e.g., $NEIGHBOUR_COUNTRIES \subset COUNTRIES \times COUNTRIES$, $STOCKS \subset PRODUCTS \times WAREHOUSES$); $\mathcal{V}$ is a non-empty collection of *value sets* (e.g., *SEXES* = {"F","M"}, $[0,16] \subset$ NAT(2), ASCII(32) $\subset$ ASCII($n$), [1/1/100, *Today*()] $\subset$ DATETIME, with NAT($n$) being the subset of naturals of at most $n$ digits, ASCII($n$) the subset of the freely generated monoid over the ASCII alphabet only including strings of maximum length $n$, etc.); $*S$ is a collection of *computed sets* (e.g., *MALES*, *FEMALES*, *UNPAID_BILLS*), and $SysS$ is a collection of *system sets* (e.g., $\emptyset$ (the empty set), NULLS (the distinguished countable set of *null values*), BOOLE = {*true*, *false*}, NAT($n$), ASCII($n$), DATETIME (the set of date and time values)).

In RDM schemata, generally, the object sets are tables, the computed sets are views (queries), the system sets are corresponding db management system (RDBMS) data types, and the value sets are their needed subsets.

The *set of mappings* (*functions*) is $\mathcal{M} = \mathcal{A} \oplus \mathcal{F} \oplus *\mathcal{M} \oplus Sys\mathcal{M}$, where $\mathcal{A} \subset \mathrm{Hom}(S - SysS \oplus \mathcal{V}, \mathcal{V})$ is the non-empty *set of attributes* (e.g., $x : PEOPLE \leftrightarrow \mathrm{NAT}(10)$, $FirstName : PEOPLE \rightarrow \mathrm{ASCII}(64)$, $BirthDate : PEOPLE \rightarrow [1/1/\text{-}6000, Today()]$, $Sex : PEOPLE \rightarrow \{$"F", "M"$\}$, $Amount : UNPAID_BILLS \rightarrow (0, 100000]$, where $\leftrightarrow$ denotes injections (one-to-one functions) and $x$ is our notation for *object identifiers*, i.e., functions to be implemented as auto-number surrogate primary keys); $\mathcal{F} \subset \mathrm{Hom}(S - SysS \oplus \mathcal{V}, S - SysS \oplus \mathcal{V})$ is the non-empty set of *structural functions* (e.g., $BirthPlace : PEOPLE \rightarrow CITIES$, $Capital : COUNTRIES \leftrightarrow CITIES$); $*\mathcal{M}$ is the set of *computed mappings* (e.g., $BirthCountry = Country \circ BirthPlace : PEOPLE \rightarrow COUNTRIES$, $Mother \bullet Father : PEOPLE \rightarrow (PEOPLE \cup \mathrm{NULLS})^2$), and $Sys\mathcal{M}$ is the set of *system mappings* (e.g., $\mathbf{1}_S$ (the unity function of a set *S*), *card*(*S*) (the cardinal of a set *S*), *Im*(*f* ) (the image of a function *f*, i.e., the set of the values it takes), *dom*(*f*) and *codom*(*f*) (the domain and codomain of *f*, respectively), *isNull*(*x*,*y*) (which returns *x* if it is not null, and *y* otherwise)).

In RDM schemata, the attributes, structural functions, and computed mappings are implemented as table or/and view (query) columns, with structural functions being foreign keys.

The *set of constraints* is $C = SC \oplus \mathcal{M}C \oplus OC \oplus SysC$, where the *set constraints* $SC$ contains both general ones (e.g., inclusion, equality, disjointness) and dyadic relation ones (e.g., reflexivity, symmetry, transitivity); the *mapping constraints* $\mathcal{M}C$ contains the general ones (e.g., totality, one-to-oneness, ontoness), dyadic-type self-map ones, general mapping products ones (e.g., minimal one-to-oneness, variable geometry keys and their subkeys, existence), dyadic-type homogeneous binary product ones, function diagram ones (e.g., commutativity, anti-commutativity, representative systems); the set $OC$ of *object constraints* contains any other closed Horn clauses not among the above, and the *system constraints* $SysC$ includes all needed mathematical constraints (e.g., $card(\emptyset) = 0$, $\emptyset \subseteq S$, $\emptyset \cap S = \emptyset$, $S \cap S = S \cup S = S$, $\forall S \in \mathcal{S}$).

Constraints are not absolute, but relative to the corresponding subuniverses: for example, constraint $C_1$ from Section 5 does not generally hold (e.g., there are several cities called "Paris" in the U.S.).

Structural E-RDs are directed graphs having as nodes any type of sets except for the value-type ones, and structural functions as edges. Our version of E-RDs [4] uses arrows (i.e., the standard mathematical convention for mappings) instead of diamonds for any binary functional relationship.

In (E)MDM, as a shorthand, the domain of attributes may be omitted if they are listed immediately after their domain set name, and one-to-one mappings use double arrows (e.g., see attribute *Country* $\leftrightarrow$ ASCII(255) from Section 5).

The *complexity of an algorithm* is denoted by $O(e)$, where $e$ is an algebraic expression and means that the algorithm never loops infinitely and ends in a number of steps that is a multiple of $e$. Generally, an algorithm is *sound* if it returns only true answers, *complete* if it accepts any valid inputs, and *optimal* (*efficient*) if it performs in the minimal number of steps possible for any input. Moreover, we say that an algorithm is *semi-optimal* when, even if it visits more than once an object, it processes any object only once.

In our context, *soundness* means that all object and computed sets are translated into same type of sets (i.e., rectangles or diamonds, respectively), attributes into ellipses, structural functions into arrows (all the above being dotted for computed objects), constraints into restrictions, and comments into informal description texts, while *completeness* means that all valid (E)MDM schemes are correspondingly translated.

In this paper, "mapping" and "function" are used interchangeably. Mapping totality is translated into RDM by a corresponding not null constraint.

## 4. The *REA*2 Algorithm

The reverse engineering pseudocode algorithm *REA*2 used by *MatBase* for translating (E)MDM schemes into E-R data models is shown in Figures 1 to 10. The dependencies between its 10 methods are presented in Figure 11.

Obviously, value-type sets are not figured in E-RDs: only object or computed ones (which are drawn in dotted lines) are.

| ***Algorithm REA2 (S, r) (Reverse engineering of (E)MDM schemas into E-R data models)*** |
|---|
| *Input*: the current (E)MDM scheme $\mathcal{M}$, and, optional, one of its non-value sets' name *S*, and a natural radius *r*<br>*Output*: a corresponding E-R data model *ER* = <*ERD*, *ERRS*, *ERID*><br>*Strategy*: |
| *if S* $\notin$ NULLS $\wedge$ *S* $\notin$ $\mathcal{M}$ *then display* "Unknown set name " & *S* & "!";<br>*else if S* $\notin$ NULLS $\wedge$ *r* $\in$ NULLS *then r* = 0; *end if*;<br>*ERD* = $\emptyset$; *ERRS* = $\emptyset$; *ERID* = $\emptyset$; // initialize *ER*<br>*if S* $\notin$ NULLS $\wedge$ *r* == 0 *then addSet*(*S*, *r*); // single E-RD<br>*else if S* $\in$ NULLS *then* // generate E-R data model for $\mathcal{M}$<br>*loop for all* non-value sets *S* $\in$ $\mathcal{M}$<br>*addSet*(*S*, *r*);<br>*end loop*;<br>*else subModel*($\mathcal{M}$, *S*, *r*); // generate E-R data sub-model<br>*end if*;<br>*end if*;<br>*end if*;<br>*End Algorithm REA*2; |

Figure 1: Algorithm *REA*2 (reverse engineering (E)MDM schemas into E-R data models)

## 5. Results: two reverse engineering examples

Here is an example of an (E)MDM schema – a subset of the *GENEALOGIES* sub-universe studied in [10, 11]:

***COUNTRIES***

*x* $\leftrightarrow$ NAT(3) total

*Country* $\leftrightarrow$ ASCII(255) total (There may not be 2 countries having same name.)

```
Method subModel(ℳ, S, r)

n = 0; computeCardinal(ℳ, n);              // n = card(Nodes(ℳ))
S_A array (text Set, natural len);         // stores sub-model sets
S_A.Set[0] = S; S_A.len[0] = 0; i = 1; j = 1; oldj = 0;
while i ≤ r ∧ j ≤ n ∧ j ≠ oldj;
   oldj = j; k = j – 1;              // starts from the latest stored set
   while k ≥ 0 ∧ S_A.len[k] == i – 1
      loop for all structural functions f defined on S_A.Set[k]
         S = codom(f); otherSet(S);
      end loop;
      loop for all structural functions f taking values from S_A.Set[k]
         S = dom(f); otherSet(S);
      end loop;
      k = k – 1;          // proceed bottom-up for current radius level
   end while;
   i = i + 1;             // increase current radius level
end while;
loop for all sets S in S_A.Set
     addSet(S, r);
end loop;
```

Figure 2: Method *subModel* of Algorithm *REA*2

```
Method addSet(S, r)

if S ∉ ERID then
     add S's description to ERID, including its maximum cardinal;
     if S is of entity-type then addEntityType(S);
     else addRelationshipType(S); end if;
     add S to ERRS, together with its maximum cardinality;
     addAttributes(S);
     addStructuralFunctions(S, r);
     loop for all constraints c ∈ ℳ, c ∉ ERRS, S involved in c
        add c to ERRS;
        add c's description to ERID;
     end loop;
end if;
```

Figure 3: Method *addSet* of Algorithm *REA*2

***CITIES***

$x \leftrightarrow$ NAT(6) total

*City* $\rightarrow$ ASCII(255) total

*Country* : *CITIES* $\rightarrow$ *COUNTRIES* total

*Capital* : *COUNTRIES* $\leftrightarrow$ *CITIES* (No city may simultaneously be the capital of 2 countries.)

$C_1$: *City* • *Country* key (There may not be 2 cities with the same name in the same country.)

$C_2$: *Country* ° *Capital* reflexive (The capital city of any country must belong to that country.)

```
Method otherSet(S)

// adds S to S_A.Set, if it is not stored in it already
if S ∉ Im(S_A.Set) then
   S_A.Set[j] = S;  // add S to the discovered sets
   S_A.len[j] = i;  // path length from center to S
   j = j + 1;       // other set found
end if;
```

Figure 4: Method *otherSet* of Algorithm *REA*2

```
Method computeCardinal(M, n)

loop for all non-value sets S ∈ M
     n = n + 1;
end loop;
```

Figure 5: Method *computeCardinal* of Algorithm *REA*2

```
Method addAttributes(S)

loop for all attributes a defined on S
   attach to rectangle or diamond labeled S an
      ellipse labeled a, dotted if a is computed;
   add to ERRS a's codomain restriction;
   addFunctRestrictions(a);
end loop;
```

Figure 6: Method *addAttributes* of Algorithm *REA*2

***DYNASTIES***

*x* ↔ NAT(8) total

*Dynasty* ↔ ASCII(255) total (There may not be 2 dynasties having same name.)

*Country* : *DYNASTIES* → *COUNTRIES* total

**_Method addStructuralFunctions(D, r)_**

*if not* (*D* is of entity-type *and* $r == 0$) *then*
*loop for all* structural functions $f: D \rightarrow C$ (including roles, if *D* is of relationship type, and canonical injections for all inclusions $D \subseteq C$), if $r \neq 0$, *or* only for *D*'s roles, if $r == 0$ *and D* is of relationship-type
*if* $C \neq D$ then *addSet*(*C*);
draw from the rectangle or diamond labeled *D* to the rectangle or diamond labeled *C* an arrow labeled *f*, dotted if *f* is computed;
*addFunctRestrictions*(*f*);
*end loop*;
*end if*;

Figure 7: Method *addStructuralFunctions* of Algorithm REA2

**_Method addFunctRestrictions(f)_**

*if f* is computed *then* add to *ERRS* and *ERID* its computation formula;
*else* add to *ERID* all information on *f*;
*if f total or f* is a role or a canonical injection *then* add to *ERRS f mandatory*;
*if f* has a default value *v then* add to *ERRS f*'s *default value* is *v*;
*if f* is one-to-one or a canonical injection *then* add to *ERRS f unique*;
*if f* is onto *then* add to *ERRS f onto*;
*if f* is a self-map *then* add to *ERRS* all *f*'s dyadic-type constraints;
*end if*;

**Figure 8: Method _addFunctRestrictions_ of Algorithm REA2**

**_Method addRelationshipType(R)_**

*if R* is a computed set *then*
add to *ERD* a dotted diamond labeled *R*;
*else* add to *ERD* a diamond labeled *R*; *end if*;

Figure 9: Method *addRelationshipType* of Algorithm *REA*2

**_Method addEntityType(E)_**

*if E* is a computed set *then*
add to *ERD* a dotted rectangle labeled *E*;
*else* add to *ERD* a rectangle labeled *E*; *end if*;

Figure 10: Method *addEntityType* of Algorithm *REA*2

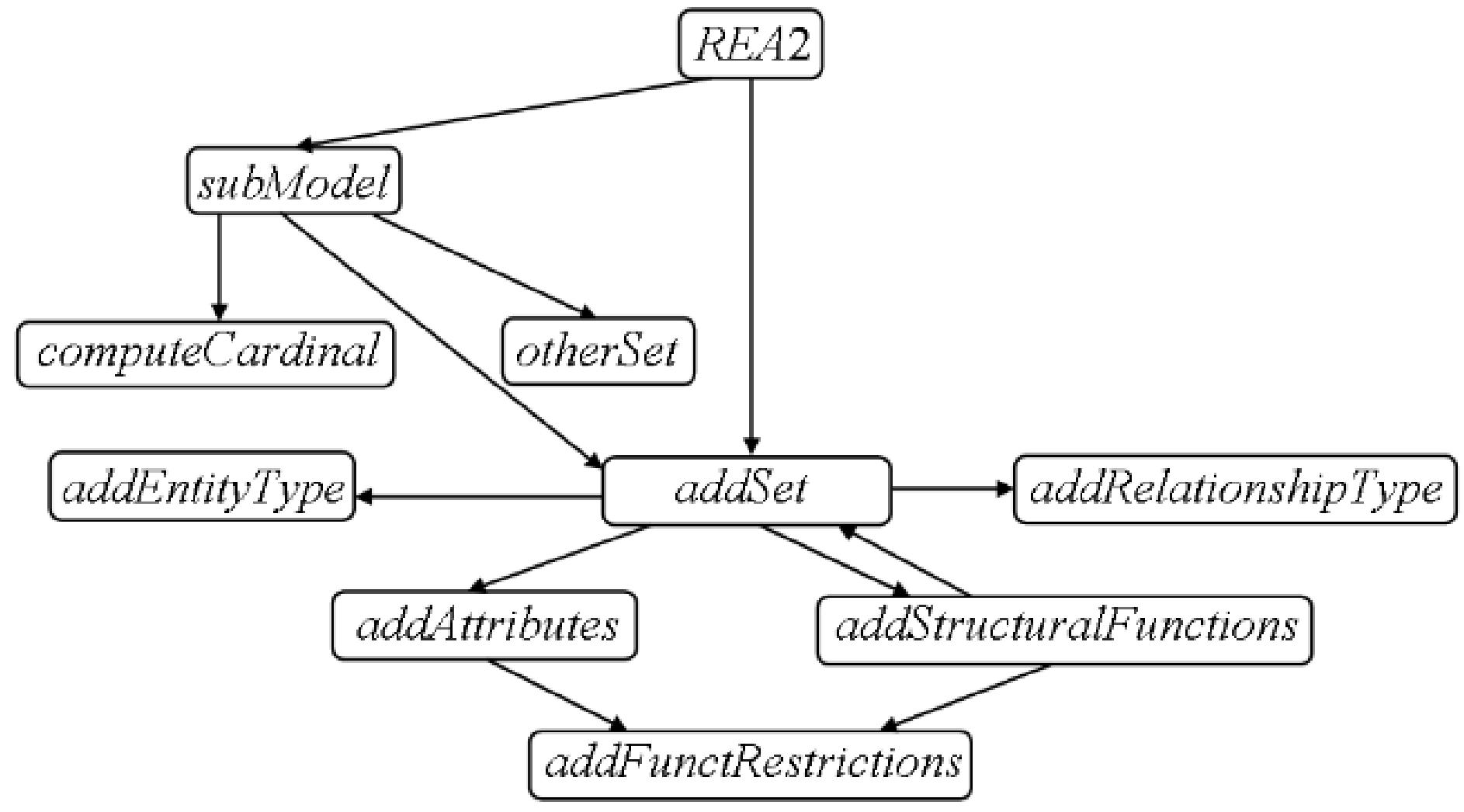

Figure 11: Algorithm *REA*2 methods' dependencies

***TITLES***

$x \leftrightarrow$ NAT(2) total

*Title* $\leftrightarrow$ ASCII(32) total (There may not be 2 titles having same name.)

***RULERS***

$x \leftrightarrow$ NAT(16) total

*Name* $\rightarrow$ ASCII(255) total

*Sex* $\rightarrow$ {'M', 'F', 'N'} total ('N' is for non-persons)

*BirthYear* $\rightarrow$ [-6500, *CurrentYear*()]

*PassedAwayYear* $\rightarrow$ [-6500, *CurrentYear*()]

*URL* $\leftrightarrow$ ASCII(255) (Any URL is dedicated to only one ruler.)

*Age* = *isNull*(*PassedAwayYear*, *CurrentYear*()) – *BirthYear*

*Mother* : *RULERS* $\rightarrow$ *RULERS* acyclic (Nobody may be his/her maternal ancestor or descendant.)

*Father* : *RULERS* $\rightarrow$ *RULERS* acyclic (Nobody may be his/her paternal ancestor or descendant.)

*KilledBy* : *RULERS* $\rightarrow$ *RULERS*

$\mathit{Dynasty} : \mathit{RULERS} \to \mathit{DYNASTIES}$

$\mathit{Title} : \mathit{RULERS} \to \mathit{TITLES}$

$\mathit{Founder} : \mathit{DYNASTIES} \leftrightarrow \mathit{RULERS}$ (Nobody founds more than one dynasty.)

$\mathit{BirthPlace} : \mathit{RULERS} \to \mathit{CITIES}$

$\mathit{Nationality} : \mathit{RULERS} \to \mathit{COUNTRIES}$

$\mathit{PassedAwayPlace} : \mathit{RULERS} \to \mathit{CITIES}$

$C_3$: $\mathit{Name} \bullet \mathit{Dynasty} \bullet \mathit{BirthYear}$ key (There may not be two persons of the same dynasty born in the same year and having the same name.)

$C_4$: $\mathit{Dynasty} \circ \mathit{Founder}$ reflexive (The founder of any dynasty must belong to that dynasty.)

$C_5$: $(\forall x \in \mathit{RULERS})(\mathit{Dynasty}(x) \notin \mathrm{NULLS} \wedge \mathit{Founder}(\mathit{Dynasty}(x)) \notin \mathrm{NULLS} \wedge \mathit{BirthYear}(\mathit{Founder}(\mathit{Dynasty}(x))) \notin \mathrm{NULLS} \Rightarrow \mathit{PassedAwayYear}(x) \in \mathrm{NULLS} \vee \mathit{PassedAwayYear}(x) > \mathit{BirthYear}(\mathit{Founder}(\mathit{Dynasty}(x))))$ (Nobody may belong to a dynasty that was founded after his/her death.)

$C_6$: $(\forall x \in \mathit{RULERS})(0 \leq \mathit{Age}(x) \leq 140)$ (Anybody's age must be a natural at most equal to 140.)

$C_7$: $(\forall x \in \mathit{RULERS})(\mathit{Sex}(\mathit{Mother}(x)) = \text{‘F’})$ (Mothers' sex must be 'F'.)

$C_8$: $(\forall x \in \mathit{RULERS})(\mathit{Sex}(\mathit{Father}(x)) = \text{‘M’})$ (Fathers' sex must be 'M'.)

$C_9$: $(\forall x \in \mathit{RULERS})(\mathit{Sex}(x) = \text{‘N’} \Rightarrow \mathit{Mother}(x) \in \mathrm{NULLS} \wedge \mathit{Father}(x) \in \mathrm{NULLS} \wedge \mathit{Dynasty}(x) \in \mathrm{NULLS} \wedge \mathit{KilledBy}(x) \in \mathrm{NULLS})$ (Non-persons may not have parents or killers or belong to dynasties.)

$C_{10}$: $\mathit{Mother} \bullet \mathit{Name}$ key (No mother gives the same name to 2 of her children.)

$C_{11}$: $\mathit{Father} \bullet \mathit{Name}$ key (No father gives the same name to 2 of his children.)

$C_{12}$: $(\forall x \in \mathit{RULERS})(\mathit{BirthYear}(x) \notin \mathrm{NULLS} \wedge \mathit{Mother}(x) \notin \mathrm{NULLS} \wedge \mathit{BirthYear}(\mathit{Mother}(x)) \notin \mathrm{NULLS} \wedge \mathit{Sex}(x) \neq \text{‘N’} \Rightarrow 5 \leq \mathit{BirthYear}(x) - \mathit{BirthYear}(\mathit{Mother}(x)) \leq 75 \wedge (\mathit{PassedAwayYear}(\mathit{Mother}(x)) \notin \mathrm{NULLS} \Rightarrow (\mathit{BirthYear}(x) \leq \mathit{PassedAwayYear}(\mathit{Mother}(x))))$

(Women may give birth only between 5 and 75 years, and before passing away.)

$C_{13}$: $(\forall x \in RULERS)(BirthYear(x) \notin \text{NULLS} \wedge Father(x) \notin \text{NULLS} \wedge BirthYear(Father(x)) \notin \text{NULLS} \wedge Sex(x) \neq \text{'N'} \Rightarrow 9 \leq BirthYear(x) - BirthYear(Father(x)) \leq 100 \wedge (PassedAwayYear(Father(x)) \notin \text{NULLS} \Rightarrow (BirthYear(x) \leq PassedAwayYear(Father(x)) + 1))$ (Men may have children only between 9 and 100 years, and at most one year after passing away.)

$C_{14}$: $(\forall x \in RULERS)(KilledBy(x) \notin \text{NULLS} \wedge Sex(x) \neq \text{'N'} \wedge BirthYear(KilledBy(x)) \notin \text{NULLS} \Rightarrow BirthYear(KilledBy(x)) \leq PassedAwayYear(x) \leq isNull(PassedAwayYear(KilledBy(x)), BirthYear(KilledBy(x)) + 140))$ (Any killer of a person must have been alive when his/her victim was killed.)

$C_{29}$: $KilledBy \vdash PassedAwayYear$ (Death year must be known for any killing.)

***MARRIAGES***

$x \leftrightarrow$ NAT(16) total

$MarriageYear \rightarrow [-6500, CurrentYear()]$

$DivorceYear \rightarrow [-6500, CurrentYear()]$

$Husband : MARRIAGES \rightarrow RULERS$ total

$Wife : MARRIAGES \rightarrow RULERS$ total

$C_{15}$: $Husband \bullet Wife \bullet MarriageYear$ key (Nobody may marry the same person twice in a same year.)

$C_{16}$: $Husband \bullet Wife \bullet DivorceYear$ key (Nobody may divorce the same person twice in a same year.)

$C_{17}$: $(\forall x \in MARRIAGES)(MarriageYear(x) \notin \text{NULLS} \Rightarrow DivorceYear(x) \notin \text{NULLS} \vee DivorceYear(x) \geq MarriageYear(x))$ (Nobody may divorce somebody before marrying him/her.)

$C_{18}$: $(\forall x \in MARRIAGES)(Sex(Wife(x)) = \text{'F'})$ (Wives' sex must be 'F'.)

$C_{19}$: $(\forall x \in MARRIAGES)(Sex(Husband(x)) =$ 'M') (Husbands' sex must be 'M'.)

$C_{20}$: $(\forall x \in MARRIAGES)(MarriageYear(x) \notin \text{NULLS} \Rightarrow ((BirthYear(Husband(x)) \in \text{NULLS} \vee BirthYear(Husband(x)) \leq MarriageYear(x)) \wedge (PassedAwayYear(Husband(x)) \in \text{NULLS} \vee PassedAwayYear(Husband(x)) \geq MarriageYear(x))) \wedge (BirthYear(Wife(x)) \in \text{NULLS} \vee BirthYear(Wife(x)) \leq MarriageYear(x)) \wedge (PassedAwayYear(Wife(x)) \in \text{NULLS} \vee PassedAwayYear(Wife(x)) \geq MarriageYear(x))))$ (Nobody may marry before being born or after death.)

$C_{21}$: $(\forall x \in MARRIAGES)(DivorceYear(x) \notin \text{NULLS} \Rightarrow ((BirthYear(Husband(x)) \in \text{NULLS} \vee BirthYear(Husband(x)) \leq DivorceYear(x)) \wedge (PassedAwayYear(Husband(x)) \in \text{NULLS} \vee PassedAwayYear(Husband(x)) \geq DivorceYear\ (x))) \wedge (BirthYear(Wife(x)) \in \text{NULLS} \vee BirthYear(Wife(x)) \leq DivorceYear(x)) \wedge (PassedAwayYear(Wife(x)) \in \text{NULLS} \vee PassedAwayYear(Wife(x)) \geq DivorceYear(x))))$ (Nobody may divorce before being born or after death.)

$C_{30}$: $(\forall x \in MARRIAGES)(PassedAwayYear(Husband(x)) \notin \text{NULLS} \wedge PassedAwayYear(Wife(x)) \notin \text{NULLS} \Rightarrow (BirthYear(Husband(x)) \in \text{NULLS} \wedge BirthYear(Wife(x)) \in \text{NULLS} \Rightarrow -140 \leq Passed\text{-}AwayYear(Husband(x)) - PassedAwayYear(Wife(x)) \leq 140) \vee BirthYear(Husband(x)) \leq BirthYear(\ Wife(x)) \leq PassedAwayYear(Husband(x)) \vee BirthYear(Wife(x)) \leq BirthYear(Husband(x)) \leq Passed\text{-}AwayYear(Wife(x)))$ (For any marriage, husband and wife must have been simultaneously alive at least one day.)

***REIGNS***

$x \leftrightarrow$ NAT(20) total

$FromY \rightarrow$ [-6500, $CurrentYear()$] total

$ToY \rightarrow$ [-6500, $CurrentYear()$]

$Ruler : REIGNS \rightarrow RULERS$ total

$Country : REIGNS \to COUNTRIES$ total

$Title : RULERS \to TITLES$

$C_{22}$: $Ruler \bullet Country \bullet FromY$ key (It does not make sense to store twice that a ruler began ruling a country during a year.)

$C_{23}$: $Ruler \bullet Country \bullet ToY$ key (It does not make sense to store twice that a ruler ended ruling a country during a year.)

$C_{24}$: $(\forall x \in REIGNS)(ToY(x) \in \text{NULLS} \vee ToY(x) \geq FromY(x))$ (No reign may end before starting.)

$C_{25}$: $(\forall x \in REIGNS)((BirthYear(Ruler(x)) \notin \text{NULLS} \Rightarrow BirthYear(Ruler(x)) \leq FromY(x)) \wedge (PassedAwayYear(Ruler(x)) \notin \text{NULLS} \Rightarrow ToY(x) \notin \text{NULLS} \wedge PassedAwayYear(Ruler(x)) \geq ToY(x))$ (Nobody may reign before being born or after death.)

$C_{26}$: $(\forall x,y \in REIGNS)(x \neq y \wedge Country(x) = Country(y) \wedge (FromY(y) \geq FromY(x) \wedge FromY(y) \leq isNull(ToY(x), CurrentYear() \vee FromY(x) \geq FromY(y) \wedge FromY(x) \leq isNull(ToY(y), CurrentYear()) \Rightarrow Sex(x) = \text{'N'} \vee Sex(y) = \text{'N'} \vee (Father(Ruler(y)) = Ruler(x)) \vee Father(Ruler(x)) = Ruler(y) \vee Mother(Ruler(y)) = Ruler(x)) \vee Mother(Ruler(x)) = Ruler(y))) \vee (\exists z \in MARRIAGES)(Husband(z) = Ruler(x) \wedge Wife(z) = Ruler(y) \vee Husband(z) = Ruler(y) \wedge Wife(z) = Ruler(x)))$ (No country may be simultaneously ruled by 2 persons, except for cases where at least one of them has sex 'N', or the two were married, or parent and child.)

Figure 12 presents the corresponding single E-RD obtained by applying algorithm *REA*2 to this (E)MDM schema with parameter values $S$ = "RULERS" and $r = 0$ (in fact, the self-maps *Mother*, *Father* and *KilledBy* are not generated for this E-RD, but for the one in Figure 13: we moved them here in this paper as Figure 13 would otherwise become too complicated).

The corresponding generated Restriction Set is the following:

a. $max(card(RULERS)) = 10^{16}$

b. *Data ranges*:

*Name*: ASCII(255)

*Sex*: {‘M’, ‘F’, ‘N’}

*BirthYear*: [-6500, *CurrentYear*()]

*PassedAwayYear*: [-6500, *CurrentYear*()]

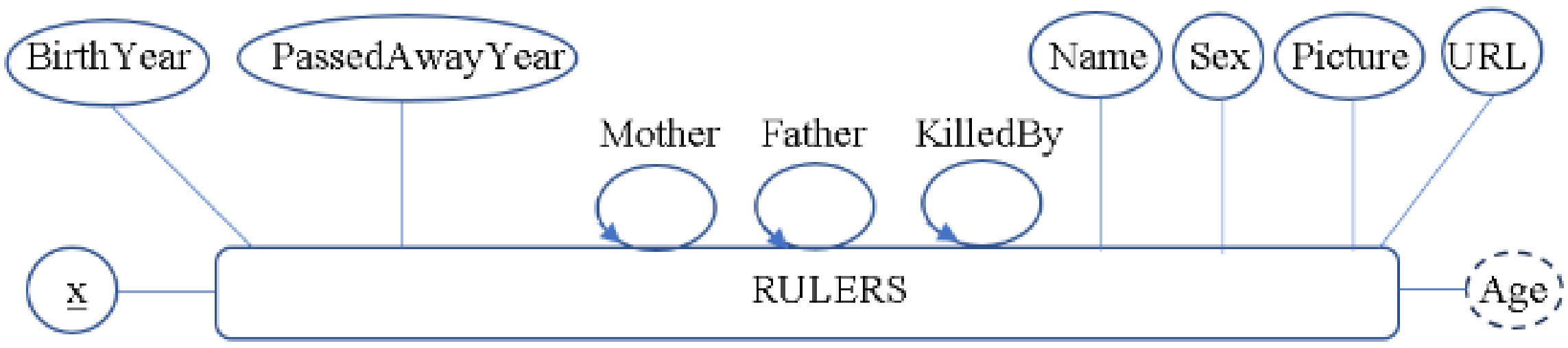

Figure 12: Single E-RD of *RULERS*

*URL*: ASCII(255)

c. *Compulsory data*: *x*, *Name*, *Sex*

d. *Uniqueness*:

*Name* • *Dynasty* • *BirthYear* (There may not be two persons of the same dynasty born in the same year and having the same name.)

*URL* (Any URL is dedicated to only one ruler.)

e. *Other types of restrictions*:

*Age* = *isNull*(*PassedAwayYear*, *CurrentYear*()) – *BirthYear*

*Mother* acyclic (Nobody may be his/her maternal ancestor or descendant.)

*Father* acyclic (Nobody may be his/her paternal ancestor or descendant.)

$C_3$ to $C_{14}$ (that we do not duplicate here, to obey paper length limitations).

The corresponding informal description is trivial: “The set of *RULERS* stores properties *x*, its object integer identifier, *Name*, an ASCII string of at most 255 characters, etc.”.

Figure 13 shows the structural E-RD obtained when running *REA*2 with *S* = “RULERS” and *r* > 0 or with null values for both *S* and *r*. The corresponding restriction set for *RULERS* is exactly the one given above; for the rest of the sets, it is similar. This goes the same for the corresponding informal description.

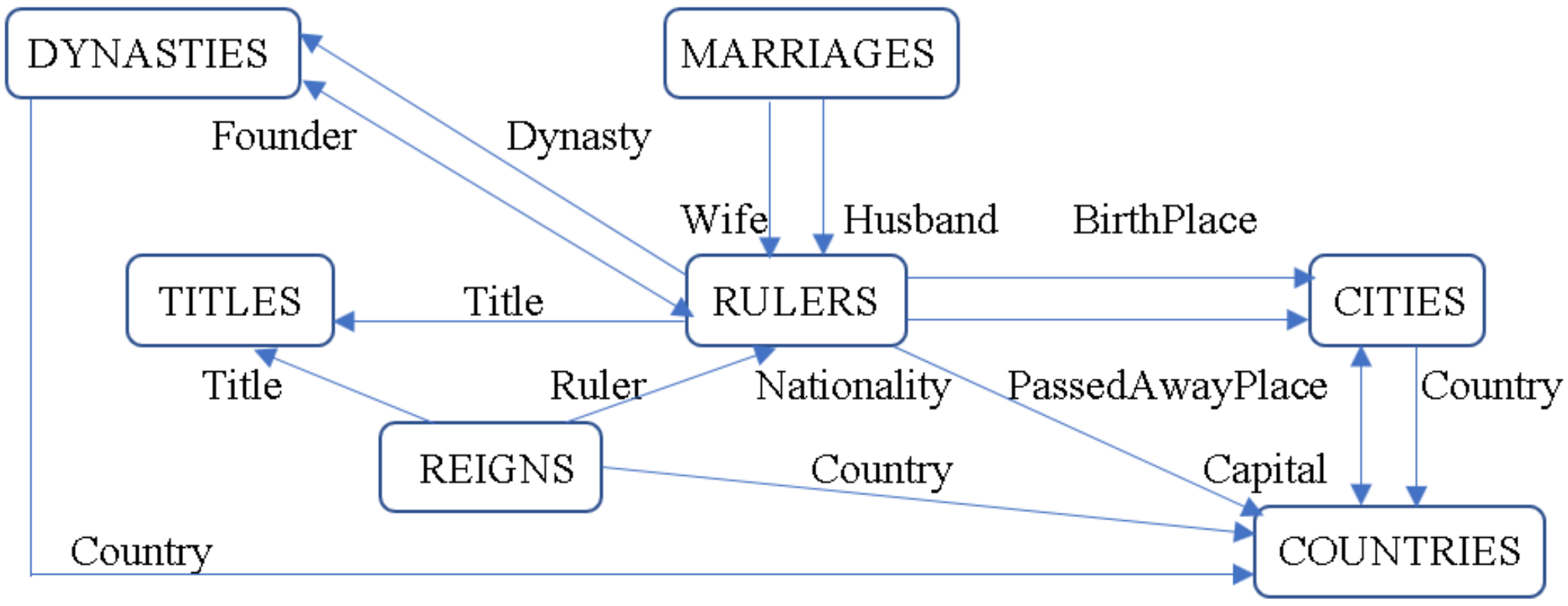

Figure 13: Structural E-RD of the (E)MDM schema

## 6. Discussion

***Proposition***

*REA*2 is:

(*i*) linear, having complexity $O(n + a + m + c)$, where $n$ is the total number of the (E)MDM schema non-value sets, $a$ is the total number of their attributes, $m$ of their structural functions (including the roles of the relationship-type sets), and $c$ is the total number of its constraints;

(*ii*) sound;

(*iii*) complete;

(*iv*) semi-optimal.

*Proof*:

(*i*) For $r = 0$, trivially, $n \leq 1$, $m = 0$, while $a$ and $c$ are the number of attributes defined on $S$ and the one of the constraints involving $S$, respectively (added by the only one executions of methods *addSet*, *addAttributes*, and *addFunctRestrictions*, when $S$ is a known set; otherwise, they are 0 as well); for $S$ known and $r > 0$, the number of the corresponding non-value-type sets of the sub-model is $s \leq n = card(Nodes(M))$, the one of structural functions is $f \leq m =$

*card*(*Edges*(M)), while *a* and *c* are the sums of the number of attributes defined on the sets of the corresponding sub-model and the sum of the constraints involving these sets, respectively; finally, when *S* is null, $s = n$, $f = m$, while *a* and *c* are the sums of the number of attributes defined on the sets of the corresponding (E)MDM schema and the sum of its constraints, respectively.

It is very easy to verify that the only loops of *REA*2 and of its method *computeCardinal* are executed exactly *n* times, the one from method *addStructuralFunctions* exactly for the number of such functions defined on its parameter, the one from *addAttributes*, similarly, exactly for the number of such functions defined on its parameter, and the one from *addSet* exactly for the number of constraints that involve its parameter and that were not previously converted to the corresponding restrictions.

The methods *otherSet*, *addFunctRestrictions*, *addRelationshipType* and *addEntityType* have no loops.

Finally, the outer *while* loop of method *subModel* is executed at most *n* times; before its first execution, the *S_A* array is initialized with the set name *S* and its level 0 (in the above example, *S* = "RULERS"); in each iteration, its inner *while* loop is executed exactly the number of times that there are sets on the previous level; in the first iteration it is executed exactly once, for *S*; its first inner loop is executed, for each iteration, exactly the number of times the current set has structural functions defined on it (e.g., for "RULERS", 5 times, discovering and storing in *S_A* the sets named "DYNASTIES", "CITIES", "COUNTRIES", and "TITLES"); its second inner loop is executed, for each iteration, exactly the number of times the current set has structural functions taking values from it (e.g., for "RULERS", 4 times, discovering and storing in *S_A* the sets named "MARRIAGES" and "REIGNS"); as, its final loop is executed at most *n* times, method *subModel* might need at most 2*n* steps to finish.

Consequently, *REA*2 never loops infinitely and is linear in the sum $n + a + m + c$.

(*ii*) As any entity-type set is translated into a rectangle, any relationship-type one into a diamond, any attribute into an ellipsis, any structural function into an arrow, any constraint into a corresponding restriction, and any comment into an informal description text, *REA*2 is sound.

(*iii*) As any well-formed (E)MDM schema (i.e., only consisting of sets, mappings, constraints, and comments) is translated into corresponding E-R data models, *REA*2 is complete.

(*iv*) For *S* null and *S* not-null but $r = 0$ or null, *REA*2 is optimal, as it reads and processes each non-value-type set, function, and constraint only once. For *S* not-null and $r > 0$, it is only semi-optimal, as it processes them only once, but reads the corresponding sub-model non-value sets twice (once for discovering and storing them into the *S_A* array, and a second time to translate them), some of the structural functions twice as well (once as defined on the current set in the inner *while* loop and a second time as taking values from it), and each constraint the number of times equal to the number of sets involved. *Q.E.D.*

As may be seen from the above example, for such small (E)MDM schemas, it is not optimal to ask for sub-models: all 7 object sets are discovered for $r = 1$, so that it would be simpler to ask for the complete model instead, as even this smallest sub-model is equal to the whole one, which can be obtained faster.

However, for commercial dbs with hundreds of object and computed sets, this feature is essential, as, on one hand, updates or/and extensions of current models are generally involving only a handful of related sets and, on the other, printing and analyzing the whole structural E-RD for them is almost impossible.

Please note that method *subModel* stops execution of its outer *while* as soon as no more sets are discovered in an iteration, which is discovered whenever at the beginning of the following one the variables *j* and *oldj* have same value. This means, for example, that running the algorithm with *S* = "RULERS" and *r* having any other natural value greater than 1 will stop exactly when the one for *r* = 1 stops.

The actual implementations of *REA*2 in *MatBase* are smarter and faster than the pseudocode algorithm presented in Section 4, by using embedded SQL over its meta-catalog. For example, in method *subModel* the *S_A* array is replaced by a temporary table with same name, and the two inner loops are replaced by the execution of the following MS VBA statement (for the structure of the meta-catalog tables *SETS* and *FUNCTIONS*, please see [8, 23]):

DoCmd.RunSQL "INSERT INTO S_A SELECT SetName, " & i & " FROM SETS INNER JOIN FUNCTIONS ON SETS.[#S] = FUNCTIONS.Codomain WHERE SetType <> 'V' AND Domain IN (SELECT [#S]  FROM SETS INNER JOIN S_A ON SETS.SetName = S_A.Set WHERE len = " & i – 1 & ") AND SetName NOT IN (SELECT Set FROM S_A) UNION SELECT SetName, " & i & " FROM SETS INNER JOIN FUNCTIONS ON SETS.[#S] = FUNCTIONS. Domain WHERE Codomain IN (SELECT [#S]  FROM SETS INNER JOIN S_A ON SETS.SetName = S_A.Set WHERE len = " & i – 1 & ") AND SetName NOT IN (SELECT Set FROM S_A)";

## 7. Conclusions and further work

To sum up, we presented a linear, sound, complete, and semi-optimal pseudocode algorithm for translating E-R data models into (E)MDM schemes, used by both versions of our intelligent DBMS prototype *MatBase*. Obviously, this algorithm may be also manually used by db and/or software architects.

We provided an example of applying it to a genealogical tree sub-universe.

We also described some powerful additional features of its actual implementations that are aimed at obtaining the fastest possible execution speed.

We successfully used this algorithm for years during our classes of Advanced Database and Software Application Architecture and Design for M.Sc. students with both Computer Science Taught in English stream of the Department of Engineering in Foreign Languages from the Bucharest Polytechnic University and Informatics of the Department of Mathematics and Informatics from the Ovidius University at Constanta, Romania.

We would warmly recommend using our (E)MDM during both db scheme design, db software application architecture, and their maintenance and extensions, as it guarantees the highest possible db data quality. We are convinced that even this small example from Section 5 proves its power and elegance, as these twin fields are, in fact, applied mathematical naïve set, relation, and function theory, plus first-order predicate calculus with equality.

Our next paper will provide the algorithm used by *MatBase* to translate (E)MDM schemas into RDM ones and associated sets of non-relational constraints –that must be enforced by the software applications managing those dbs–, which is the next step towards guaranteeing the highest possible db stored data quality. For the last step, i.e., the architecture and design of db software applications, we would warmly recommend our db constraint-driven approach [11] .

Acknowledgement

Nobody other than its authors contributed to this paper.

Funding Support

This work was not sponsored by anybody.

Ethical Statement

This study does not contain any studies with human, or animal subjects performed by any of the authors.

## Conflicts of Interest

The authors declare that they have no conflicts of interest to this work.